\documentstyle[titlepage,12pt]{article}
\begin{document}
\parindent 2em
\baselineskip 4.5ex
\begin{titlepage}
\begin{center}
\vspace{12mm}

{\LARGE First order melting of vortex lattice in
strongly type-II three dimensional superconductors}
\vspace{25mm}

Igor F. Herbut and Zlatko Te\v sanovi\' c \\
Department of Physics and Astronomy, The Johns Hopkins University,
Baltimore, Maryland 21218

\end{center}
\vspace{10mm}

\noindent
{\bf Abstract}: We calculate the transition line of the first-order
melting of vortex lattice
in a three-dimensional type-II superconductor in fields of several Tesla,
using the results from the density-functional theory of vortex melting
in two-dimensions and a self-consistent Hartree
treatment of correlations along the field. The result is in
quantitative agreement with experiment. The temperature width
of the hysteresis, the latent heat, the Debye-Waller factor
and the magnetization at the transition are discussed.


\end{titlepage}


Several recent experimental studies focused on the superconducting transition
in untwinned YBCO samples in fields of several tesla \cite{1,2,3,4}.
In contrast to
earlier experiments in samples where the physics of the transition
was dominated by
inhomogeneities and where the transition
appeared to be second order, the observed sharp hysteretic drop in
 resistivity in these very clean, strongly type-II materials
in moderate magnetic fields suggests that the true vortex melting
transition might be first order.
This is somewhat surprising when one recalls that the mean-field Abrikosov
theory predicts a second order phase transition
for a homogeneous type-II superconductor
in magnetic field. Thus here one encounters another possible example
of strong thermal fluctuations changing the order of transition. An early
suggestion that this might happen in a type-II superconductor in vicinity
of $H_{c2}(T)$
came from the renormalization group analysis \cite{5} close to the
upper critical dimension $d_{up}=6$, as well as from the analysis of the
theory with infinite number of order parameter components in $4<d<6$ \cite{6}.
This is
inadequate however for the physical
three-dimensional (3D) samples which are
below the lower critical dimension $d_{low}=4$
in the problem considered there.  Usually,
the transition in the vortex system is described by the harmonic theory
of vortex lattice and by invoking phenomenological Lindemman criterion
to locate the melting point \cite{7,8}.
This is however a suspect starting point if
one is interested in describing strong
fluctuations  near $H_{c2}(T)$ and is
more appropriate for the low-field or $T\approx 0$
region of the phase diagram.
As  emphasized by Moore \cite{9},
in this description one starts from the
Abrikosov lattice solution for the order parameter which is unstable
with respect to harmonic shear modes of the
lattice at any finite temperature
in two and three dimensions.
It is therefore inconsistent
to simply assume this ordered low-temperature
state. Furthermore, the numerical
constant in Lindemman
criterion needs to be chosen phenomenologically, and the requisite
number can actually differ by orders of magnitude from one material to
another \cite{10}.

Recently, a novel approach to the problem has been formulated that encompasses
the difficulties mentioned above by relying on a new physical picture
of the phase transition in strongly type-II system in high field \cite{11}.
Unlike the renormalization group study
of ref. 5 where the superconducting Abrikosov transition arises
from the growing phase correlations in the
directions along the magnetic field, here
the positional correlations orthogonal to the
field drive the transition. The right paradigm
is the two-dimensional (2D)
problem. In
high magnetic field (to be specified later),
the original Ginzburg-Landau partition function in the symmetric gauge
is equivalent to a system of classical
particles interacting with long-range, multi-body forces \cite{11}. It is
a scale invariant, incompressible system which undergoes a weak
first-order freezing transition, not unlike one-component Coulomb plasma in
2D. The point of transition, the latent heat
 and other relevant quantities at the transition
can be quite accurately determined by using the density-functional theory
of solidification \cite{12}. The low-temperature phase is
not the familiar Abrikosov vortex lattice
but a charge-density-wave (CDW) of Cooper
pairs, with a weak periodic modulation of Cooper
pair density but with
no long-range phase coherence \cite{11,12a}. In this paper we
extend the theory to 3D superconductor
and use it
to calculate the phase boundary in $H-T$
phase diagram.  We find a very good quantitative
agreement with experiment. Also, we make a number of
predictions about the typical physical quantities at the transition
and analyze the temperature width of the hysteresis as a function of the
field.

Consider the Ginzburg-Landau partition function for an
anisotropic, homogeneous, 3D superconductor in strong magnetic field,
with fluctuations of the magnetic field neglected (Ginzburg-Landau
parameter $\kappa >>1$).
For the fields above $H_{b}\approx (\theta/16)H_{c2}(0)(T/T_{c0})$,
where $\theta=\Lambda_{eff}/\Lambda_{T}$
($\Lambda_{eff}=2\lambda_{ab}(0)^{2}/d$, $\Lambda_{T}=\phi_{0}^{2}/
16\pi^{2}T_{c0}$) is the Ginzburg fluctuation parameter
the fluctuation spectrum is dominated by the lowest Landau level
modes (LLL) \cite{13}. In YBCO $\theta\sim 0.04$  and $H_{c2}(0)\sim
150T$ so the approximation which
retains only the LLL modes should be appropriate for the fields larger
than $\sim 0.3T$. The partition function is $Z=\int D[\psi^{*},
\psi] exp(-S)$, and
\begin{equation}
S=\frac{1}{T}\int d^{2}\vec{r} dz [\gamma |\partial_{z}\psi|^{2} + \alpha'(T,H)
|\psi|^{2}+\frac{\beta}{2}|\psi|^{4}]
\end{equation}
where $\alpha'(T)=\alpha(T)(1-H/H_{c2}(T))$, and $\alpha(T)=a(T-T_{c0})$,
$\beta$
and $\gamma$ are phenomenological parameters. It is convenient to
rescale the fields and lengths as $(2d\beta2\pi l^{2}/T)^{1/4} \psi
\rightarrow\psi$, $r/(l\sqrt{2\pi})\rightarrow r$ and $z/d\rightarrow z$,
where $l$ is the magnetic length for charge $2e$,
and $d$ is a microscopic length
along the field (typically the spacing between the pairs of CuO planes).
The action in the exponent then becomes
\begin{equation}
S=\int d^{2}\vec{r} dz [g_{\gamma}|\partial_{z}\psi|^{2} +
g_{\alpha}|\psi|^{2} + \frac{1}{4} |\psi|^{4}],
\end{equation}
 with $\psi$ which is restricted to the LLL,
and the whole thermodynamics is determined by two
dimensionless coupling constants
$g_{\gamma/\alpha}= \{\gamma/d^{2},\alpha\}\times\sqrt{\pi l^{2} d/\beta T}$.
In the 2D case, $g_{\gamma}=0$, and the system described by the
partition function (2) undergoes a weak
first-order transition into CDW phase of Cooper pairs at $g_{\alpha}
\equiv g_{M}=-6.5$, as
seen in Monte-Carlo simulations \cite{11,14,15,15a} and found in
density-functional
theory \cite{12}. The CDW has a triangular modulation with the
period set by the magnetic length: $a=l\sqrt{4\pi/\sqrt{3}}$.
Small deviations from this periodicity are expected in principle,
but do not change our results substantially and will not be considered here.
We assume that the vortex transition in 3D system is driven by the same
mechanism of growing positional correlations between vortices.
In 2D, $g_{\alpha}\propto d^{1/2}$, where $d$ is the film thickness.
In 3D, this length is replaced by some temperature and
field-dependent length $\Lambda$ over which the
vortices are ``straight" along the field direction. $\Lambda$ provides
a short length scale along the $z$-axis, just like the magnetic length
does in the $x-y$ plane.  Here we
assume that this length is not very different from the
superconducting correlation length
along the field, $\xi_{||}$. Thus
we make an ansatz that there is a single coupling constant which
describes the physics of the transition in the 3D regime (i. e., for
$\xi_{||}>1$ in units of $d$): $g_{\alpha}\xi_{||}^{1/2}$. The phase boundary
is then determined by the condition
\begin{equation}
g_{M}= g_{\alpha} \xi_{||}^{1/2},
\end{equation}
for $\xi_{||}>1$.
To explicitly determine the transition line in the $H-T$ phase
diagram we use the correlation length as obtained in the
self-consistent Hartree treatment of the theory defined by eq. 2:
\begin{equation}
\xi_{||}=(\frac{g_{\gamma}}{g_{\alpha}+<|\psi|^{2}>/4})^{1/2}
\end{equation}
where the thermal average appearing is determined by the equation
\begin{equation}
<|\psi|^{2}>=\frac{1}{2\pi}
\int_{-\infty}^{\infty} \frac{dk}{g_{\gamma}k^{2}+g_{\alpha}+<|\psi|^{2}>/4}.
\end{equation}
The expressions 4 and 5 look the same as they would in a purely 1D
theory. This is a consequence of the strong magnetic field, which
causes $D\rightarrow D-2$ dimensional reduction in the theory on the
Hartree level.  The full theory as given in eq. 2 is, of course, not 1D,
due to the non-local constraint on the fields to be entirely in the LLL.
In fact, the transition we are describing comes exactly form those lateral,
intra-LLL
correlations which would be undetectable in a simple Hartree theory.
$\xi_{||}$, however, is expected to be reasonably
well described by the Hartree theory,
since the dimensional reduction is exact for $\langle\psi^{*}\psi\rangle$
correlator \cite{11,15b}.

{}From the melting condition 3 and using the expressions 4 and 5
one obtains the equation for the
transition line
\begin{equation}
t+h+(\frac{2c\kappa^{2}\xi_{ab}^{2}(0)}{\Lambda_{T}\xi_{c}(0)}
)^{2/3} (t h)^{2/3}=1,
\end{equation}
where we rescaled the field and the temperature as $t=T/T_{c0}$,
$h=H/H_{c2}(0)$, and the constant $c$ is
determined by the numerical value of $g_{M}$  as
$c^{2}=g_{M}^{4} (\sqrt{1+ 1/g_{M}^{2}}-1)/2 =10.5$. The equation
for the transition line 6 is our main result. It depends only on a
particular combination of the material constants and on a pure number
 $g_{M}$.
Its accuracy is restricted to the region
where $\xi_{||}>1$ since otherwise the layers would decouple and the
description in terms of 2D melting would be appropriate.
In the region of 3D to 2D crossover better results would be
obtained by using the Lawrence-Doniach model. We chose anisotropic
Ginzburg-Landau partition function instead because the transition line can be
found in simple closed form, which still is accurate over a large
portion of the phase diagram.

In figure 1. we compare the calculated transition line with the
experimental results on superconducting transition in YBCO
in the magnetic field parallel to the c-axis \cite{3}.
We set $T_{c0}=90K$, $H_{c2}(0)=148.14T$ (as found by linear
extrapolation from small fields), $\kappa=50$,
$\xi_{c}(0)=3\AA$,
and the best fit is obtained with $\xi_{ab}(0)/\xi_{c}(0)=5.8$
which agrees well with what is known about this material
\cite{16}. The results of experiments in refs. 1 and 2 are
essentially the same and are not shown. Note that the
distance between CuO planes
$d=12\AA$ completely cancels out in the determination of the
transition line.  According to our calculation, YBCO is three
dimensional everywhere along the transition line in the range
of the fields studied. One may also expect that the numerical
value of $g_{M}$ should be renormalized from its 2D value that we
used due to the difference between the lengths $\Lambda$ and $\xi_{||}$.
This question can be settled
by knowing more precisely the  values of
parameters $\kappa$ and anisotropy in eq. 6. The present
calculation suggests however that this difference is not substantial.

The density-functional theory predicts that the CDW phase can be superheated
up to $g_{SH}=-6.25$ in the 2D vortex system. If we
neglect the supercooling of the vortex liquid
upon lowering the temperature and take only the superheating
of the solid phase into account we may calculate the
thermodynamic hysteresis width
in temperature as a function of the magnetic field for 3D transition.
The result is shown on
Figure 2. The functional dependence of the calculated
hysteresis width agrees with
the  observation in resistivity measurements in
refs. 2,3,4 but the result is roughly an order of magnitude larger.
We attribute this to the effect of disorder in the
experiment, which tends
to reduce the width of hysteresis.
However, for a given sample and a small current, the hysteresis widths in
temperature at different fields, when multiplied  by a suitably chosen
fudge factor, agree well with our curve (see Figure 2.). The reason for
this is the following: in a homogeneous 2D sample, the hysteresis
width $\Delta t_{hyst}(h)\propto \Delta g= g_{SH}-g_{M}$, for
small $\Delta g$. A weak point disorder cuts off the crystalline
order at large but finite Larkin-Ovchinikov length $\xi_{LO}$. For a
crystallite of that size
both $g_{M}$ and $g_{SH}$ increase proportionally to
$\xi_{LO}^{-1}$, but their difference $\Delta g$ decreases. Thus the
primary effect of weak disorder is to decrease the numerical value
of $\Delta g$ from its thermodynamic value of 0.25 \cite{12},
and therefore decrease $\Delta t_{hyst}(h)$. The functional
dependence of the temperature width of the hysteresis on the field
however is not affected and hence the above behavior.

Although the transition under consideration is into
a 3D phase of periodically modulated density of Cooper pairs, we expect
the relevant quantities at the transition not to be very different
from 2D case. This comes as a consequence of the same mechanism
of the transition which is effectively 2D in nature.
Thus, the latent heat per vortex and per layer should be $\sim 0.3k_{B}T_{M}$
\cite{12}. Debay-Waller factor $\nu(\vec{G})=|\rho(\vec{G})|^{2}$
at the transition should be
given by $\rho(\vec{G})=0.72\exp(-\lambda^{2}G^{2})$ with $\lambda=0.47$,
where $\vec{G}$ is a reciprocal lattice vector in units where $l=1$.
The magnetization is expected to have a discontinuity of roughly 1\%
of its value at the transition \cite{12}.
The solid phase below
the transition is expected to have at most a power law superconducting
order \cite{9,11}, but it will behave as a superconductor
for most practical purposes--For example, its ohmic resistivity
will be extremely low. This is the consequence of YBCO
being effectively an anisotropic 3D superconductor in the region
of fields and temperatures where the experimental data are currently
available. This leads to an enhanced density modulation of the SCDW
below the transition with the resulting large increase in the
range of superconducting correlations. At even higher fields,
where the layered structure
of YBCO becomes more pronounced, we expect the range of superconducting
correlations to become substantially shorter, perhaps leading to
a clearly identifiable SCDW state with a long-range 3D positional
order but only a finite resistivity. In this respect, a large
variety of the HTS materials which are more anisotropic than YBCO
appear particularly promising.

Finally, we should stress that our results leave open a possibility
that $g_M (3D)$ is significantly different from $g_M (2D)$, with the
corresponding modifications in the latent heat, characteristic
vortex bending length, magnetization discontinuity,
etc. The problem is that, while our theory successfully captures
lateral correlations of the `microscopic' GL-LLL theory which are responsible
for the SCDW transition both in 2D and 3D, it still requires
as an outside input the vortex structure factor near the transition.
In 2D we were able to obtain this needed information from the
numerical Monte Carlo simulations of the GL-LLL theory. Unfortunately,
such numerical results are not available at present for the
3D or layered GL-LLL model.

In conclusion, we calculated the transition line of first order
vortex lattice melting in strongly type-II superconductor in
magnetic field of several Tesla. The agreement with the experimental
results on untwinned YBCO is excellent in the whole range of fields
of $0.1-10T$. The temperature width of the hysteresis, the latent heat,
the Debye-Waller factor at the transition and the discontinuity of
the magnetization are discussed. Our results strongly support the
conclusion that the vortex-liquid to solid transition in
homogeneous strongly type-II superconductor is first order.

This work has been supported in part by the NSF Grant No. DMR-9415549.

\pagebreak
Figure Captions:

Figure 1. H-T phase diagram for YBCO ($t=T/T_{c0}$).
The full line represents
the melting line calculated from eq. 6 with parameters
described in the text. Crosses are the experimental points from ref. 3.
The dashed line is the mean-field $H_{c2}(T)$ line.

Figure 2. The field dependence of temperature width of the
hysteresis due to superheating of the low-temperature phase.
The full line represents the calculated hysteresis width. Triangles
and crosses are experimental points of ref. 2 for two samples with
different degrees of disorder.
The temperature widths are multiplied by
factors 5 (triangles) and 10 (crosses) to fit the curve.

\pagebreak

\end{document}